\documentclass[12pt]{iopart}

\usepackage{graphicx}
\usepackage{dcolumn}
\usepackage{bm}

\begin{document}


\newcommand{\La}{La$_{1.85}$Sr$_{0.15}$CuO$_4$}
\newcommand{\Bi}{Bi$_2$Sr$_2$CaCu$_2$O$_{8+\delta}$}

\title[Critical state stability in La$_{1.85}$Sr$_{0.15}$CuO$_4$]{Influence of crystal anisotropy on the critical state stability and flux jumps dynamics in a single crystal of La$_{1.85}$Sr$_{0.15}$CuO$_4$}

\author{A Nabia{\l}ek$^1$, A Wi\'{s}niewski$^1$, V V Chabanenko$^2$, S V Vasiliev$^2$, T V Tsvetkov$^2$ and F P\'erez-Rodr\'{\i}guez$^3$}
\address{$^1$Institute of Physics, Polish Academy of Sciences, Al. Lotnik\'{o}w 32/46, 02-668 Warsaw, Poland.}
\address{$^2$Institute for Physics and Engineering, NASU, 72 ul. R. Luxemburg, 83114, Donetsk, Ukraine.}
\address{$^3$Instituto de F\'{\i}sica, Benem\'erita Universidad Aut\'onoma de Puebla, Apdo. Post. J-48 Puebla, 72570 Pue., Mexico.}
\ead{nabia@ifpan.edu.pl}
%

%
%
%
%

%
\date{\today} 

\begin{abstract}

We studied the critical state stability in a large cubic sample of a single crystalline \La \ for different sample orientations with respect 
to the external magnetic field as well as the dynamics of the flux jumps. It is shown that thermomagnetic avalanches develop in dynamic conditions 
characterized by significantly lower magnetic diffusivity than the thermal one.
In this case, critical state stability depends strongly on cooling conditions. 
We compared predictions of the isothermal model and of the model for the weakly cooled sample with experimental results.
In both models, the field of the first flux jump decreases with an increase of sweep rate of the external magnetic field.
We also investigated the influence of external magnetic field on the dynamics of the following stages of the thermomagnetic avalanche.
It is shown that the dynamics of the flux jumps is correlated with the magnetic diffusivity proportional to the flux flow resistivity. 

\end{abstract}

\pacs{74.25.Ha, 74.25.Sv}
\submitto{\SUST}

\maketitle

\section{INTRODUCTION}

At certain conditions, the critical state of a type-II superconductor may become unstable \cite{Mints81}, then thermomagnetic avalanche may develop.
During the thermomagnetic avalanche, temperature of the superconducting sample increases rapidly and large amount of the magnetic flux enters into the sample volume.
Thermomagnetic avalanches, called flux jumps, are problematic from a viewpoint of practical applications of superconductors,
 because they may drive a superconducting sample into a normal or, at least, into a resistive state.

The stability of critical state depends strongly on a relation between the thermal ($D_\mathrm{t}$) and magnetic ($D_\mathrm{m}$) diffusivity. 
$D_\mathrm{t}=\frac{\kappa}{c_\mathrm{V}}$, where $\kappa$  is the thermal conductivity, and $c_\mathrm{V}$ is the specific heat at constant volume.
$D_\mathrm{m}=\frac{\rho}{\mu_0}$.
In the flux flow regime, $\rho=\rho_\mathrm{ff}=\rho_\mathrm{n}(\frac{H}{H_\mathrm{c2}})$, where $\rho_\mathrm{ff}$ is the flux flow resistivity, $\rho_\mathrm{n}$ is the normal state resistivity,
$H_\mathrm{c2}$ is the second critical field and $\mu_0$ is the magnetic permeability of vacuum. In the case of hard conventional superconductors, usually $\tau=\frac{D_\mathrm{t}}{D_\mathrm{m}}<<1$.
Such conditions are called "locally adiabatic". The most commonly studied parameter, which determines the stability of the critical state is a field of the first (after cooling the sample in zero magnetic field) flux jump, $H_\mathrm{fj1}$.
This parameter is closely correlated with the critical penetration depth, $L_\mathrm{c}$. If the critical current density, $j_\mathrm{c}$, does not depend on the magnetic field, $L_\mathrm{c}=\frac{H_\mathrm{fj1}}{j_\mathrm{c}}$.
This parameter is very important from a viewpoint of the practical applications of the superconductors, because it is possible to avoid the thermomagnetic avalanches, if the width of the superconducting sample is smaller than $2L_\mathrm{c}$.
We can also express the condition of the stability of the critical state in a form: $H_\mathrm{fj1}>H_\mathrm{p}$, where $H_\mathrm{p}$ is the field of full penetration,
$H_\mathrm{p}=j_\mathrm{c}a$, and $a$ is half-width of the investigated sample. 
In the simplest approximations, the sample is usually assumed to be of a shape of an infinite slab, of width $2a$, and an external magnetic field is applied parallel to its surface.
In locally adiabatic conditions, the field of the first flux jump can be expressed by a formula

\begin{eqnarray}
 H_\mathrm{fj1} &=& \sqrt{ \frac {2c_\mathrm{V}j_{c}} {\mu_0 |\frac{\partial j_{c}}{\partial T}|} }
\label{eq:one}
\end{eqnarray}

The thermal parameter, which determines the critical state stability in locally adiabatic conditions, is the specific heat $c_\mathrm{V}$.
In locally adiabatic conditions, the stability of the critical state is not influenced by the cooling conditions at the surface of the superconducting sample.

The situation is different in a case when $\tau >> 1$. Such conditions usually occur in superconducting composites and are called "dynamic".
In dynamic conditions, the thermal parameter which determines the stability of the critical state is the thermal conductivity, $\kappa$, or the thermal boundary conductivity, $h$ \cite{Swarz89}.  

Some problems connected with the critical state stability in high temperature superconductors (HTSC) are not clarified yet.
The problems are connected with the complex structure of these materials.
The analysis of the critical state stability in HTSC is usually performed within approximations developed for conventional superconductors \cite{Mints81}, which sometimes lead to wrong results.
Because of the increasing number of applications of HTSC, it is very important to understand all aspects of the thermomagnetic avalanche development in these materials.
One of the specific features of HTSC is their strong crystallographic anisotropy.

Experimental data show that high temperature superconductors are usually more stable against flux jumping than conventional superconductors.
The larger stability of the high temperature superconductors against flux jumping can be correlated with the flux creep phenomenon, which is usually strong in these materials  \cite{McHenry92, Gerber93}.  
In studies of the magnetization of superconductors, flux creep manifests itself as a relaxation of the magnetic moment.
The process of magnetic relaxation reflects strongly nonlinear current-voltage characteristics of superconducting sample.
These characteristics can be approximated by the formula

\begin{eqnarray}
 j(E) &=& j_\mathrm{c}+\frac{j_\mathrm{c}}{n} \ln \left( \frac{E}{E_0} \right),
\label{eq:two}
\end{eqnarray}

\noindent where $E_0$ is the voltage criterion at which the critical current density, $j_\mathrm{c}$, is defined, and $n$ is a dimensionless parameter.
One usually assumes $E_0=10^{-4}$ V/m. In this case $n>>1$.

The influence of nonlinear current-voltage characteristics on the critical state stability, in the case of a weakly cooled sample (for $Bi=\frac{ah}{\kappa} \ll 1$,
$Bi$ is so-called Biot number), was analyzed in Ref.\cite{Mints96}. 
It was shown that in such conditions the critical state stability depends on the external magnetic sweep rate,$\frac{\partial{H_\mathrm{ext}}}{\partial{t}}$.

\begin{eqnarray}
 H_\mathrm{fj1}^\mathrm{w} &=& j_\mathrm{c}\sqrt{\frac{2\mu_0 h}{n\left|\frac{\partial j_\mathrm{c}}{\partial T}\right| \frac{\partial H_\mathrm{ext}}{\partial t}}}
\label{eq:three}
\end{eqnarray} 

In previous works \cite{Nab03, Nab06}, we studied the influence of external magnetic sweep rate on the critical state stability in crystals of high temperature \Bi \ superconductor.
It is extremely difficult to obtain thick crystals of \Bi, and we have shown \cite{Nab06} that the thickness of the crystal strongly influences the critical state stability.
On the other hand, it is relatively easy to obtain large cubic samples of \La.
For this reason, we have chosen this system to study the influence of crystal anisotropy on the critical state stability.

A strong influence of cooling condition and a role of crystal anisotropy on the critical state stability in the case of textured YBaCuO were considered in Ref. \cite{Guillot, Watanabe}.
The effects of thermal insulation on the critical state stability in conventional NbTi superconductor were investigated in Ref. \cite{Chikaba}.

We paid special attention to the dynamics of the flux jumps, which remains poorly understood up to date.
This dynamics is also important from a viewpoint of superconductor applications, because a large power of the heat released during short avalanches may, at certain conditions, lead to the destruction of superconducting devices.
In the present work, we also studied the dynamics of the magnetic flux changes in a single crystal of \La \ during thermomagnetic avalanches.
In particular, we focused on the influence of external magnetic field on dynamics of the following stages of the thermomagnetic avalanche.

\section{EXPERIMENT}

In our investigations, we used a large 3 x 3 x 3 $\mathrm{mm}^3$ single crystal of \La, obtained by the floating zone technique.
In order to study the influence of the external magnetic sweep rate on the critical state stability, we used the Quantum Design PPMS system with the extraction magnetometer option and the maximal external magnetic field attainable of 9 tesla.
After cooling the sample in zero magnetic field, external magnetic field was swept with constant rate from 0 to 9~T.
In the following experiments, we varied the sweep rate from the minimal ($10^{-3}$ T/s) to the maximal ($2 \cdot 10^{-2}$ T/s) one attainable in our system.
The distance between the following data points taken in each magnetization curve was limited by the sweep rate and the time needed to take each data point.
At the maximal sweep rate ($2 \cdot 10^{-2}$ T/s), we were not able to obtain the distance smaller than about 0.1 T.
This distance has limited an accuracy of $H_\mathrm{fj1}$, determined in our experiments.

In order to determine the current-voltage characteristics of our sample, we performed studies of magnetic moment relaxation.
After cooling in zero external magnetic field, external magnetic field was first swept, with minimal sweep rate, up to 0.5 T,
 and next the relaxation of magnetic moment was registered at constant external magnetic field during approximately 1 hour.

The studies of $H_\mathrm{fj1}$, as well as of magnetic relaxation, were performed both for external magnetic field parallel to the $c$-axis and for external magnetic field parallel to the $ab$-plane. 

The dynamics of the flux jumps was studied in the Cryogenics 12 tesla magnet system with variable temperature insert for external magnetic field parallel to the $c$-axis.
A pick-up coil consisted of 6 turns of copper wire wound around the investigated sample and connected to data acquisition board (DAQ) in the computer.  
Because of small inductance of the applied coil and large internal resistance of DAQ, we were able to register the changes of the voltage with time resolution better that 10$^{-6}$~s.
The time resolution of DAQ was about 10$^{-7}$~s.
We registered time dependence of the coil voltage during the following flux jumps.
Additionally, we used two miniature Hall probes to register magnetic field dependence of the surface self-field, $H_\mathrm{self-surf}$, of the investigated sample.
One of the probes was put in the centre of the surface of the investigated sample.
The second probe measured external magnetic field. The differential signal from the two probes was proportional to $H_\mathrm{self-surf}$.
The cryogenic Hall probes were made from tin doped InSb films.
The probe was supplied with the current of 10 mA.
The sensitivity of the Hall probe was about 5 mV/T.

The rate of external magnetic field sweep was approximately 1 T/min. In the 12 tesla magnet system, we also performed magnetization measurements using vibrating sample magnetometer (VSM).

It is important to note that the heat exchange conditions in both systems used in our experiments were very different.
In the PPMS system, the sample was put into a standard PPMS holder, which was surrounded by helium gas with low pressure (about 0.5 Torr).
In the 12 tesla magnet system, during the experiments at 2 K and 4.2 K, the sample was immersed in liquid helium.
During experiments at higher temperatures, the sample was surrounded by flowing helium gas with approximately atmospheric pressure.
The sample was either glued to a quartz plate (studies of the flux jumps dynamics) or put into Teflon holder (VSM).

\section{RESULTS}

\begin{figure}[t] \centering
\includegraphics[width=0.95\columnwidth]{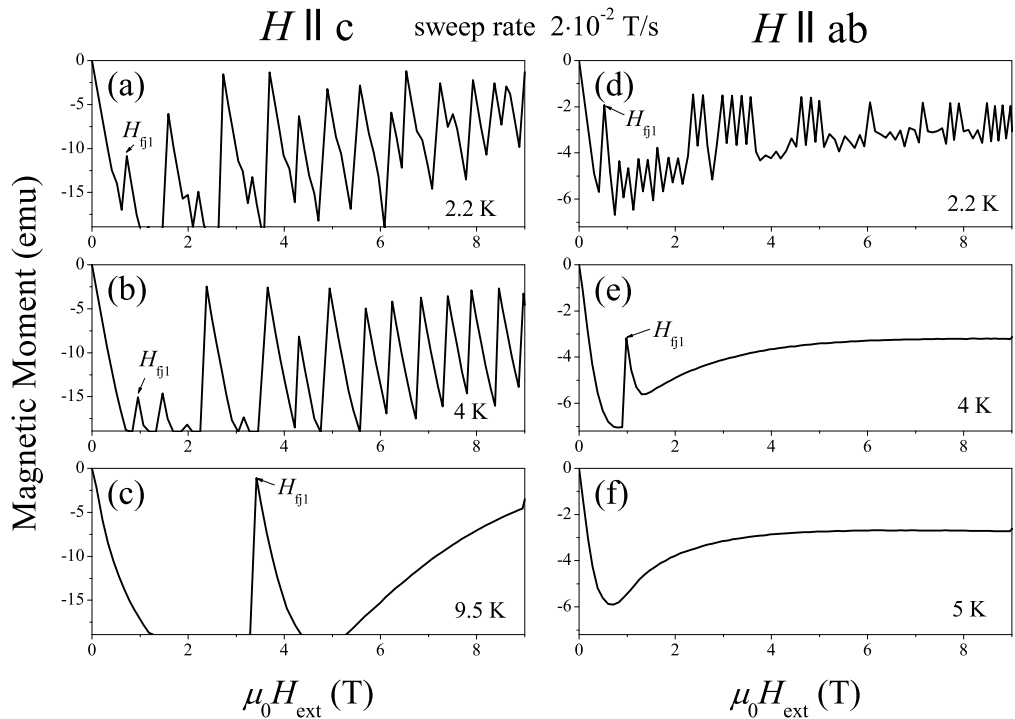}
\caption {Virgin magnetization curves taken by extraction magnetometer (PPMS) with the sweep rate of $2 \cdot 10^{-2}$ T/s at different temperatures for both sample orientations studied in our experiments.}
\end{figure}

\begin{figure}[t] \centering
\includegraphics[width=0.95\columnwidth]{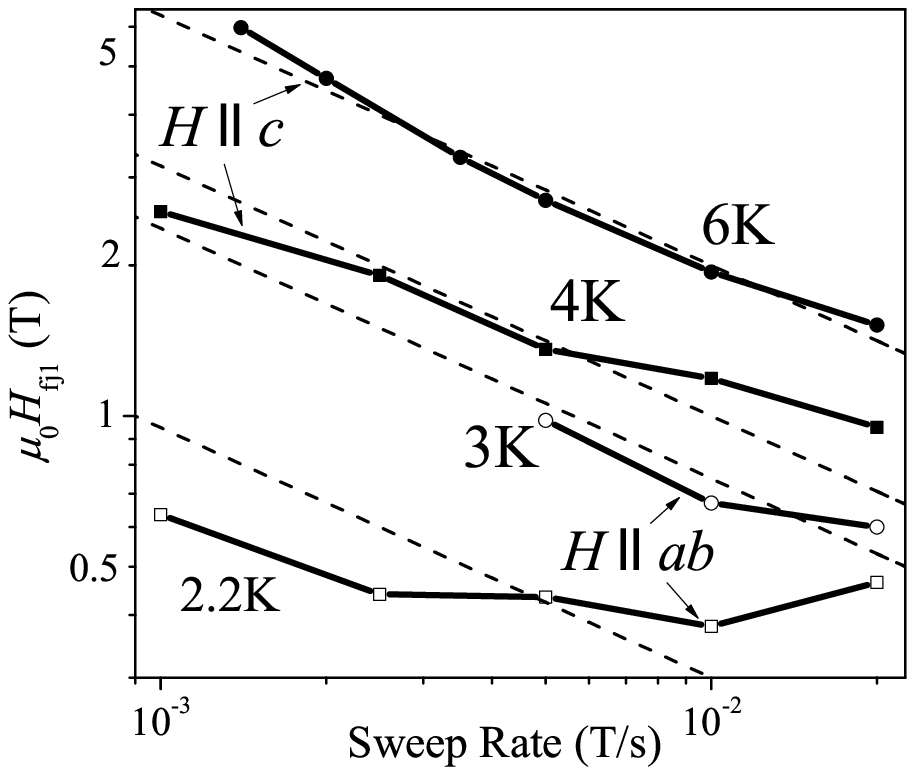}
\caption {The changes of the field of the first flux jump, $H_\mathrm{fj1}$, with the sweep rate at different temperatures and for different sample orientations
 with respect to the external magnetic field, in a double-logarithmic scale, for the 3 x 3 x 3 $\mathrm{mm}^3$ \La \ single crystal surrounded by helium gas (PPMS experiments).
 Dashed lines show the predictions of Eq. 3, where $H_\mathrm{fj1}$ is inversely proportional to the square root of the sweep rate.  }
\end{figure}

The range of flux jumps occurrence, the field of the first flux jump as well as the distance between the following flux jumps depends on the sample orientation, temperature, sweep rate and the experimental system.
Figure 1 presents some of the virgin magnetization curves taken by extraction magnetometer (PPMS) with the sweep rate of $2 \cdot 10^{-2}$~T/s at different temperatures for the both sample orientations studied in our experiments.
Magnetization curves on the left side in Fig.~1 are cut-off for the signal amplitude above 20 emu, because of a limitation of the signal in extraction magnetometer.
Nevertheless, in all our PPMS experiments, we were able to determine the value of $H_\mathrm{fj1}$. 
Temperature range of the flux jumps occurence depends on the sample orientation.
For the external magnetic field parallel to the $c$-axis, flux jumps occur even at 9.5 K (see Fig. 1c).
For the external magnetic field parallel to the $ab$-palne, flux jumps disappear already at 5 K (see Fig. 1f).
With the increase of temperature, we observed an increase of $H_\mathrm{fj1}$ and of the distance between the following flux jumps.
We also observed a decrease of $H_\mathrm{fj1}$ with an increase of external magnetic field sweep rate (see. Fig. 2)
Hence, we observed the flux jumps in the largest temperature range, for the highest sweep rate ($2 \cdot 10^{-2}$ T/s).
At the lowest sweep rate, $10^{-3}$ T/s, we did not observe any flux jumps at the temperatures higher than 4 K, for the external magnetic field parallel to the $c$-axis,
and higher than 2.2 K, for the external magnetic field parallel to the $ab$-plane, respectively.
Figure 2 shows the changes of $H_\mathrm{fj1}$ with the sweep rate at different temperatures and for different sample orientations - the data were taken from PPMS experiments.
The results are shown in a double-logarithmic scale.
If $H_\mathrm{fj1}$ changes with the sweep rate according to the power law, we expect a straight line in such plot.
Dashed lines show the predictions of Eq. 3, where $H_\mathrm{fj1}$ is inversely proportional to the square root of the sweep rate.

The experiments in the 12 tesla magnet (with Hall probes or VSM) were all performed with the external magnetic field sweep rate approximately the same as the highest sweep rate in the PPMS system.
However, we observed here the flux jumps only at the lowest temperature of about 2.2 K and for the external magnetic field parallel to the $c$-axis (see Fig. 3a - experiment with the Hall probes).
No jumps were present neither at 4.2 K nor at higher temperatures. 

\begin{figure}[t] \centering
\includegraphics[width=0.95\columnwidth]{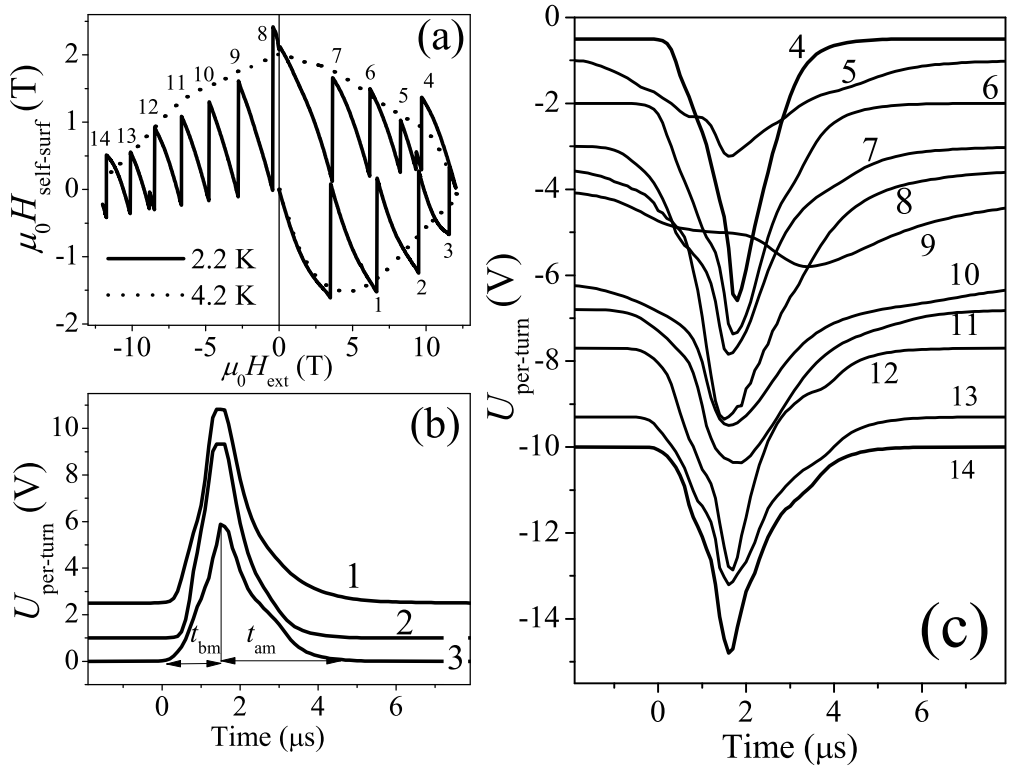}
\caption {a) Hysteresis loops of the surface self-field, $H_\mathrm{self-surf}$, measured by the Hall probes in 12 tesla magnet system at 2.2 K and 4.2 K - sample immersed in liquid helium.
 b,c) Time dependence of the voltage, per one turn of the coil, registered during the following flux jumps.
 The numbers correspond to the numbers denoting successive flux jumps presented in part (a). Initial voltages of the following flux jumps were shifted, in order to present all data in one figure.
 $t_\mathrm{bm}$ and $t_\mathrm{am}$ describe duration of the two characteristic stages of the flux jump. }
\end{figure}

The relaxation of the magnetic moment, measured in external magnetic field $\mu_0 H_\mathrm{ext}=0.5$ T, was approximately logarithmic in time,
 which was consistent with the model of thermally activated motion of bundles of vortices \cite{Anderson}.

In Figure 3a, on can see the hysteresis loops of the surface self-field, $H_\mathrm{self-surf}$, measured by the Hall probes in 12 tesla magnet system
 at the temperature of about 2.2 K (solid line) and 4.2 K (dotted line), for the external magnetic field parallel to the $c$-axis.
At 2.2 K, one can see giant jumps of the surface self-field caused by the thermomagnetic avalanches.
No jumps were present at the temperature of 4.2 K or at higher temperatures.

In Figure 3b and 3c, one can see time dependence of the voltage per one turn of the coil wounded around the investigated sample,
registered during the succeeding flux jumps.
The numbers at the curves correspond to the numbers denoting successive flux jumps presented in Fig. 3a.
Typical duration of the flux jumps was of several microseconds.
The dependence shown in Fig. 3b corresponds to the lower (ascending) branch of the hysteresis loop.
The dependence shown in Fig. 3c corresponds to the upper (descending) branch of the hysteresis loop, shown in Fig. 3a.
Before and after each flux jump, the voltage of the coil was close to zero.
However, in order to present all the data in one figure, the initial voltages of the following flux jumps were shifted.
The voltage of the coil is proportional to the time derivative of the magnetic flux in the investigated sample.
We can recognize two characteristic stages of the flux jump.
During the first stage, the rate of the magnetic flux changes increases.
During the second stage, the rate of the magnetic flux changes decreases.
We have denoted duration of these two stages as $t_\mathrm{bm}$ ("before maximum") and $t_\mathrm{am}$ ("after maximum"), respectively (see Fig. 3b).  
We have determined duration of the following stages of the flux jump at a noise level, which in our system was about 30 mV.
One can see that the voltage of the observed impulses for the ascending branch of the hysteresis loop (Fig. 3b) has the sign opposite to the descending branch (Fig. 3c).
This is because, for the ascending branch, the magnetic flux enters into the investigated sample during the succeeding flux jumps,
 while for the descending branch, we observe flux exit.

\section{ANALYSIS}

\subsection{The current-voltage characteristics}

In order to compare the models of the critical state stability with our experimental results, it is necessary to know
the current-voltage characteristics of the investigated sample \cite{Mints96}.
These characteristics can be determined using the data of the magnetic moment relaxation.
In order to obtain these characteristics, we applied the following procedure.
We used the model of an infinite slab sample.
Additionally, we assumed the screening current density to be the same in the whole sample volume penetrated by the magnetic flux.

The time dependent screening current density can be in this case correlated with the magnetization.
The electric field at the surface of the slab is proportional to the time derivative of the screening current density.
Treating the time as a parameter, one can obtain the current-voltage dependence, $j(E)$.

The obtained by the described above procedure current-voltage characteristics, were fitted according to Eq. 2.
We applied this procedure for both sample orientations studied in our experiments.
Our fitting procedure gave the critical current density of an order of $10^8$ A/$\mathrm{m}^2$ and $10^9$ A/$\mathrm{m}^2$
 and the $n$-parameter (see Eq. 2) of about 40 and 30 for the external magnetic field parallel to the $ab$-plane and parallel to the $c$-axis, respectively. 
In the range of the flux jumps occurrence, the $n$-parameter was only weakly dependent on temperature. Hence, we did not consider this dependence in our further analysis.

The temperature dependence of the critical current density was estimated using the widths of the magnetization hysteresis loops in the external magnetic field of 1 T,
 which was close to the field of the first flux jump, and the formulas from Ref. \cite{Gyorgy}.
We used here the loops taken by VSM in 12 tesla magnet system.
In VSM system, we did not observe any flux jumps neither at 4.2 K nor at higher temperatures, and there was no limitation of the signal above 20 emu like in PPMS extraction magnetometer.
Limitation of the signal as well as a large number of the flux jumps observed in PPMS experiments made determination of the critical current density (using PPMS data) impossible.

The experimentally obtained temperature dependence of the critical current density was fitted according to the exponential formula: $j_\mathrm{c}(T)=j_\mathrm{c0} \exp \left( -\frac{T^{*}}{T_0} \right)$,
where $T^{*}=\frac{T}{g(T)}$, $g(T)=1- \left( \frac{T}{T_\mathrm{c}} \right)^2 $ and $T_\mathrm{c}=~$35.5~K.
We have found the following fitting parameters.
For the external magnetic field parallel to the $c$-axis, $j_\mathrm{c0} = 5.6 \cdot 10^9$ A/$\mathrm{m}^2$ and $T_0$~=~7.4~K 
(the critical current density in the direction parallel to the $ab$-plane).
For the external magnetic field parallel to the $ab$-plane, $j_\mathrm{c0} = 7.3 \cdot 10^8$ A/$\mathrm{m}^2$ and $T_0$~=~4.3~K 
(the critical current density in the direction perpendicular to the $ab$-plane).
Our fitting procedure was performed only in the range 4 - 15 K, which was most interesting from the viewpoint of the flux jumps occurrence.

\subsection{The $\tau$-parameter and the Biot number}

In order to determine the conditions of the critical state stability and the thermomagnetic avalanche development,
 it is necessary to find the relation between the thermal and magnetic diffusivity or the parameter $\tau = \frac{D_\mathrm{t}}{D_\mathrm{m}} = \mu_0 \frac{\kappa}{c_\mathrm{V}} \sigma $.
In the case of the flux creep model: $\sigma = \frac{j_\mathrm{c}}{n E}$.
In the case of an infinite slab model, the electric field at the surface of the slab: $E=\mu_0 a \left( \frac{H}{H_\mathrm{p}} \right) \frac{\partial H_\mathrm{ext}}{\partial t}$
for $H<H_\mathrm{p}$ and $E=\mu_0 a \frac{\partial H_\mathrm{ext}}{\partial t}$ for $H>H_\mathrm{p}$.
The maximal value of the electric field which can be, at given conditions, induced in superconducting sample $E_\mathrm{max}=\mu_0 a \frac{\partial H_\mathrm{ext}}{\partial t}$.
Using the maximal value of the electric field we can calculate the minimal value of $\tau$, $\tau_\mathrm{min}=\frac{\kappa j_\mathrm{c}}{{c_\mathrm{V}} n a \left( \frac{\partial H_\mathrm{ext}}{\partial t} \right)}$.
In our analysis, we used the temperature dependence of the thermal parameters $\kappa$ and $c_\mathrm{V}$ from Ref. \cite{Morelli} and Ref. \cite{Chen}, respectively.

In all our calculations, we used the value of in-plane thermal conductivity, which is approximately one order of magnitude higher than out-of-plane thermal conductivity \cite{Morelli}.
As a result, the ability of investigated sample to remove thermal fluctuations is limited by the in-plane thermal conductivity.

We have found that in the case of all our experiments $\tau>\tau_\mathrm{min}>10^2>>1$.
Hence, we can assume the dynamic conditions of the thermomagnetic avalanche development.

In dynamic conditions, we expect the critical state stability to be dependent on the cooling conditions.
This is consistent with the results of our experiments.
In order to analyze this problem, we calculated the Biot number, $Bi=\frac{a \ h}{\kappa}$.
The Biot number depends on the thermal boundary conductivity, $h$.
This parameter is very difficult to evaluate experimentally \cite{Swarz89}.
For the sample immersed in liquid helium at normal pressure and in the absence of the boiling crisis, (12 tesla magnet experiments), we expect $h$ to be of an order of $10^4$ W/$\mathrm{m}^2$K \cite{Mints81}.

In the case of PPMS experiments, we expect this parameter to be of several orders of magnitude lower.
The criteria of the critical state stability can be easily derived in two limiting cases.
1) $Bi<<1$ - weakly cooled sample (this case was considered in Ref. \cite{Mints96} ),
2) $Bi>>1$ - so-called isothermal conditions.

We have found that the approximation of the weakly cooled sample cannot be applied in the case of our sample immersed in liquid helium (in this case $Bi\sim$ 10).
However, it seems to be reasonable to use this approximation in the case of PPMS experiments, where we expect $Bi$ to be several orders of magnitude lower. 

\subsection{The weakly cooled sample - PPMS experiment}

We assumed the dynamic model for the weakly cooled sample ($Bi<<1$) to describe the critical state stability in the case of our experiments performed in PPMS.
Hence, we used Eq. 3, and our experimental data, to calculate the thermal boundary conductivity, $h$.
If the boundary conductivity is governed by phonon processes, we expect this dependence
 to be described by a power function with the power of about 3 \cite{Swarz89}.
Hence, we assumed $h=c \cdot T^3$.
For the highest sweep rate $2 \cdot 10^{-2}$ T/s, we have found, from the fit to experimental data, $c=0.25$ W/$\mathrm{m}^2$K$^4$ or $c=2.8$ W/$\mathrm{m}^2$K$^4$
 for the external magnetic field parallel to the $c$-axis or to the $ab$-plane, respectively.  
At 4.2~K the calculated, according to our procedure, thermal boundary conductivity was 2-3 orders of magnitude lower
 than that for the sample immersed in liquid helium at normal pressure.
Thus our assumption about a low value of the Biot number seems to be correct.

However, it is necessary to notice that Eq.~3 does not describe precisely the experimentally observed dependence of $H_\mathrm{fj1}$ on the sweep rate (see Fig.~2).
If $H_\mathrm{fj1}$ changes with the sweep rate according to Eq.~3, we expect the dashed lines in Fig.~2 to be parallel to the experimental curves.
We observed the larges disagreement for the external magnetic field parallel to the $ab$-plane.

\subsection{The critical state stability and flux creep in the isothermal approximation}

In order to derive the critical state stability criterion in the isothermal conditions ($Bi>>1$), we used the following approximations.
Let us assume the infinite slab to be penetrated by the magnetic flux to the depth $d$.
To simplify further calculations, we assume the space coordinate $x=d$ at the sample border.
According to the arguments presented in Ref. \cite{Mints96} in dynamic conditions for an infinite slab model,
 we can correlate the fluctuation of the electric field  $\delta E(x)$ with the fluctuation of the temperature  $\delta T(x)$ according to the formula:

\begin{eqnarray}
 \delta E(x)=\frac{n E(x)}{j_\mathrm{c}} \left| \frac{\partial j_\mathrm{c}}{\partial T} \right| \delta T(x)
\label{eq:ten}
\end{eqnarray}

The background electric field, $E(x)$, is induced by the sweep of the external magnetic field: $E(x)=\mu_0 \frac{\partial H_\mathrm{ext}}{\partial t} x$.
In the case of weakly cooled sample ($Bi<<1$), one can assume the thermal fluctuation to be independent on the space coordinate.
The stability criterion (Eq. 3) can be in this case derived from the inequality $\int_0^d j \delta E dx \leq h \delta T(d) $, 
where the left side of the inequality is the heat generated by the fluctuation in the sample volume and the right side is the heat removed from the sample through its border \cite{Mints96}. 

In isothermal conditions ($Bi>>1$), because of strong cooling of the sample, we can assume the fluctuation of the temperature
 at the sample border to be equal to zero,  $\delta T(d)=0$.
In our approximation, we assumed the space distribution of the thermal fluctuation to be described by a square function $\delta T(x)=\delta T_0 \left( 1-\frac{1}{d^2}x^2 \right)$
where  $\delta T_0=\delta T(x=0)$ is the maximal temperature fluctuation at the penetration depth.
The maximal heat power, generated in the sample by the fluctuation, which can be removed from the sample volume,
 is in this case limited by the thermal conductivity, and the critical state stability criterion can be described by the inequality
$j\delta E(x)\leq -\kappa \frac{\partial^2}{\partial x^2} \delta T(x)$.
Hence, $j \delta E(x)=n \mu_0 \frac{\partial H_\mathrm{ext}}{\partial t} \left| \frac{\partial j_\mathrm{c}}{\partial T} \right|\delta T_0 x \left( 1-\frac{1}{d^2} x^2 \right)\leq - \kappa \frac{\partial^2}{\partial x^2} \left( \delta T_0 \left( 1-\frac{1}{d^2} x^2 \right) \right) = \frac{2 \kappa}{d^2} \delta T_0$.

The left side of this inequality has a maximum for $x=\frac{d}{\sqrt{3}}$. 
As a result, the critical state stability criterion is given by the condition $d\leq L_\mathrm{c} = \sqrt[3]{\frac{\gamma \mu_0 \kappa}{n \left| \frac{\partial j_\mathrm{c}}{\partial T} \right| \frac{\partial H_\mathrm{ext}}{\partial t}}}$, or 

\begin{eqnarray}
H_\mathrm{fj1}^{\mathrm{is}} = j_\mathrm{c} \sqrt[3]{\frac{\gamma \mu_0 \kappa}{n \left| \frac{\partial j_\mathrm{c}}{\partial T} \right| \frac{\partial H_\mathrm{ext}}{\partial t}}}
\label{eq:eleven}
\end{eqnarray}

\noindent where $\gamma = \frac{3}{13} \left( 9+ \sqrt{3} \right) \approx 2.48$.
It is important to note that, also in the case of isothermal approximation,
 we expect the field of the first flux jump to decrease with an increase of the sweep rate, $ \frac{\partial H_\mathrm{ext}}{\partial t}$.

\subsection{Comparison of theoretical models with the experimental results}

\begin{figure}[t] \centering
\includegraphics[width=0.95\columnwidth]{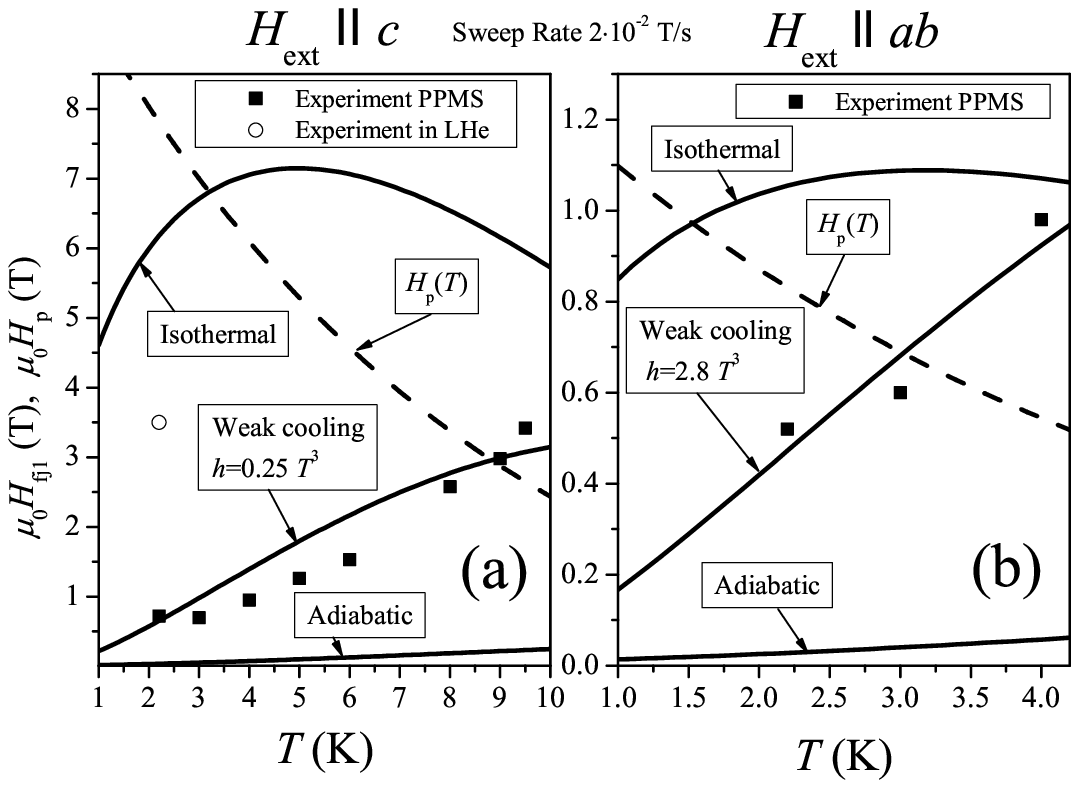}
\caption {Temperature dependence of the field of the first flux jump, $H_\mathrm{fj1}$ at the sweep rate of $2 \cdot 10^{-2}$ T/s and in the external magnetic
 field parallel to the $c$-axis (a) and parallel to the $ab$-plane (b) of the 3 x 3 x 3 $\mathrm{mm}^3$ single crystal of \La.
 The predictions of the approximated models of the critical state stability - see Eqs. 1, 3 and 5 are compared with the experimental data. Dashed line shows temperature dependence of the field of full penetration, $H_\mathrm{p}(T)$. }
\end{figure}

In Figure 4, we present the comparison of the predictions of the three theoretical approximations discussed in the present work,
 i.e. the local adiabatic approximation and two dynamic approximations (of weakly cooled sample and isothermal)
 with the experimentally obtained temperature dependence of $H_\mathrm{fj1}$.
The experimental data were taken for both sample orientations studied in our experiments: in the external magnetic 
field parallel to the $c$-axis (Fig. 4a) or parallel to the $ab$-plane (Fig. 4b), taken at the sweep rate of $2 \cdot 10^{-2}$~T/s.
For the external magnetic field parallel to the $c$-axis, we present the results both in the case when the sample is cooled by gaseous He
 (the PPMS experiment) and in the case when the sample is cooled by liquid He (the experiment in 12 tesla magnet with the Hall probe - see Fig.~3a, in this case $\mu_0H_\mathrm{fj1}$(2.2~K)~=~3.5~T).
In the both figures, we also present temperature dependence of the field of full penetration: $H_\mathrm{p} (T)$.
This parameter is very important, because in the infinite slab model of the critical state stability, we expect the flux jumps to vanish when $H_\mathrm{fj1}$ exceeds $H_\mathrm{p}$.
For the model of weak cooling, we assumed $h=c \cdot T^3$, where $c=0.25$~W/$\mathrm{m}^2$K$^4$ or $c=2.8$~W/$\mathrm{m}^2$K$^4$
 for the external magnetic field parallel to the $c$-axis or to the $ab$-plane, respectively.
It is clearly seen that our experimental results cannot be explained by the local adiabatic approximation.
The best fit of the PPMS results was found in the dynamic model of weak cooling,
 while the results obtained with the 12 tesla magnet are better described by the isothermal approximation.
It is worth to note that the isothermal curve in Fig.~4a crosses the $H_\mathrm{p}(T)$ curve at the temperature of about 3 K.
This result can explain the fact that, in the case of 12 tesla magnet experiment,
 we observed the flux jumps only at the temperature of 2 K and not at temperatures higher than 4.2 K.
On the other hand, in the case of PPMS experiments, we observed the flux jumps also above the $H_\mathrm{p}(T)$ curve.
It is worth to note that also in other experiments, one sometimes observes flux jumps in the external magnetic field higher
 than the field of full penetration \cite{Gerber00}.
Such behavior of the flux jumps is not fully clarified yet.

\subsection{Dynamics of the flux jumps}

\begin{figure}[t] \centering
\includegraphics[width=0.95\columnwidth]{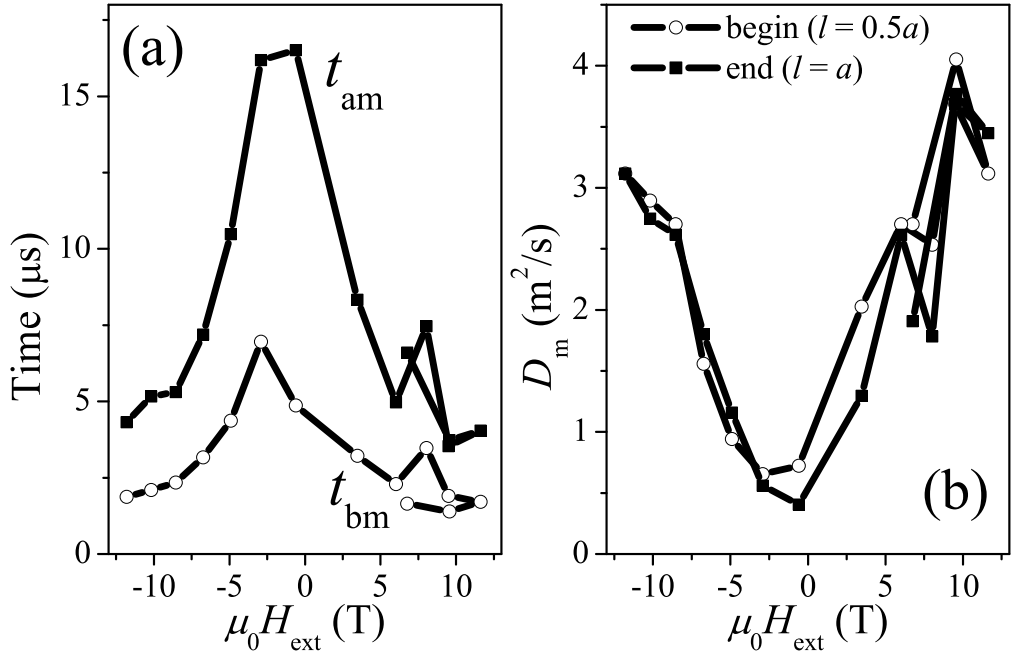}
\caption {a) The magnetic field dependence of the characteristic times $t_\mathrm{bm}$ and $t_\mathrm{am}$, before and after maximum of the coil voltage.
 b) The magnetic field dependence of the magnetic diffusivity, estimated form the exponential fit of the coil voltage at the begin
 and at the end of the flux jumps, respectively. }
\end{figure}

Duration of the two characteristic stages of the flux jumps denoted in Fig. 3b as $t_\mathrm{bm}$ and $t_\mathrm{am}$
 decreases with an increase of the external magnetic field, and $t_\mathrm{bm}$ was approximately three times shorter than $t_\mathrm{am}$ (see Fig. 5a).

The exponential character of the final stage of the flux jump suggests that it can be analyzed in terms of the magnetic diffusion.
The process of the magnetic diffusion depends on the initial magnetic field distribution in the sample.
It is possible to decompose this distribution into a series of functions characteristic for the sample geometry
 (e.g. a series of cosines in the case of infinite slab geometry or a series of Bessel functions in the case of a sample with cylindrical symmetry).
Each harmonic of this decomposition has specific time dependence.
From the sum of the time dependent harmonics, one can obtain the time dependent distribution of the magnetic flux in the investigated sample.
The voltage of the coil, wound up around the investigated sample, is proportional to the time derivative of the magnetic flux.
We assumed the sample to be of a shape of infinite slab.

For sufficiently long times in comparison to the parameter

\begin{eqnarray}
t_0 = \frac{4 l^2}{\pi^2 D_\mathrm{m}},
\label{eq:fourteen}
\end{eqnarray}

\noindent we can assume the voltage of the coil - $U(t)\sim \exp(-t/t_0)$.

The parameter $l$ is a characteristic length of the diffusion process. If the slab is fully penetrated by magnetic flux, $l=a$.
Hence, when one fits the time dependence of the coil voltage at the final stage of the flux jump, one can find the coefficient $D_\mathrm{m}$.
The magnetic field dependence of this parameter is shown in Fig. 5b.
We have found an increase of this parameter with increasing external magnetic field.

The initial exponential increase of the voltage also depends on the magnetic diffusivity of the superconducting sample \cite{Mints81}.
We have fitted this dependence by an exponential function $(U(t)\sim \exp(t/t_0)$).
However, it is necessary to discuss the meaning of the length parameter (the parameter $l$ in Eq. 6) at the begin and at the end of the flux jump.
During the flux jumps observed in our experiment, the sample was wholly penetrated by the magnetic flux (see Fig. 3a).
Hence, as a characteristic length of the diffusion process (in Eq. 6), at the end of the flux jump, we assumed the parameter $a$,
 it means half of the sample width.
On the other hand, at the begin of the flux jump, only a part of the sample is penetrated by the magnetic flux.
Hence at this stage, one should assume the characteristic length parameter to be equal to the penetration depth, which is smaller than $a$.
The magnetic field dependence of the parameter $D_\mathrm{m}$ for the begin of the flux jump,
 we have found to be similar as that for the end of the jump.
These two parameters coincide very well, if we assume in Eq.~6 \ $l=0.5 a$ (see Fig. 5b) for the begin of the flux jump.

The characteristic diffusivity found in our experiment was of an order of 1 $\mathrm{m}^2$/s (see Fig. 5b).
Such an order of diffusivity is characteristic for the magnetic diffusion in our system,
 providing the sample is in the normal state or in the flux flow regime.
The resistivity of \La \ is anisotropic. However, we studied dynamics of the flux jumps only for the external magnetic field parallel to the $c$-axis of the investigated crystal.
In such sample orientation, screening currents flow only in the $ab$-plane.
If we assume the in-plane normal state resistivity 
 $\rho_\mathrm{n}> 0.25 \ \mathrm{m} \Omega \mathrm{cm}$ \cite{Fuji}, then $D_\mathrm{m} = \rho_\mathrm{n} / \mu_0 > 2 \ \mathrm{m}^2/\mathrm{s} $.
On the other hand, if one takes data of the thermal conductivity ($\kappa$) \cite{Morelli} and of the specific heat ($c_\mathrm{V}$) \cite{Chen},
 one can estimate the in-plane thermal diffusivity ($D_\mathrm{th}=\kappa/c_\mathrm{V}$) to be of an order of $10^{-3} - 10^{-2} \ \mathrm{m}^2/\mathrm{s}$.
The out-of-plane thermal diffusivity is approximately one order of magnitude lower than the in-plane thermal diffusivity \cite{Morelli}.  
 
\section{Discussion} 

The critical state stability in a single crystalline \La \ is influenced by a large number of parameters.
All of them must be considered in order to determine the critical state stability in the framework of existing theoretical models.
In \La \ crystal, most of these parameters are anisotropic.
Hence, the stability of the critical state depends strongly on the sample orientation with respect to the external magnetic field.

In our analysis, we consider both anisotropy of critical current density and the anisotropy of the current-voltage characteristic (the anisotropy of the $n$-parameter).
In the present analysis, we assumed the thermal conductivity of the investigated sample to be limited only by the in-plane thermal conductivity,
 which is approximately one order of magnitude higher that the out-of-plane thermal conductivity.
According to our results in the case of \La \ crystal, the anisotropy of the $n$-parameter is relatively weak.
Hence, the anisotropy of the critical state stability is connected mainly with the anisotropy of screening currents.
The difference between the in-plane and out-of-plane screening current, at given temperature, is about one order of magnitude.

In both sample orientations studied in our experiments, the thermomagnetic avalanche develops in dynamic conditions.
Our sample is characterized by a strong flux creep phenomenon.
Hence, the critical state stability is influenced both by the external magnetic field sweep rate and the cooling conditions.
These cooling conditions were very different in two cryostats used in our experiments.

In PPMS experiments, where the sample is surrounded by helium gas of low pressure,
 the critical state stability can be analyzed in the approximation of weak cooling or low Biot number.
In such condition, the thermal parameter that influences the critical state stability is the thermal boundary conductivity.
This parameter is difficult to determine experimentally.
It depends e.g. on the roughness of the surface of the investigated sample.
In the case of PPMS experiments, the investigated sample is put additionally into a holder
 which makes an analysis of the heat exchange conditions even more difficult.
Nevertheless, it seems to be slightly surprising that, according to our analysis, 
the thermal boundary conductivity changes by an order of magnitude after the change of the sample orientation.

It is possible that such results are, to some extent, connected with the approximations assumed by the models used in our analysis.
In our analysis we used models of an isotropic infinite slab sample. 
In order to determine (in the framework of these models) the critical state stability criterion, it is necessary to know a large number of parameters.
Each of these parameters must be determined experimentally and all parameters have some experimental errors.
As the result, the uncertainty of the obtained critical state stability criterion can be relatively large.
The models used in our analysis can be applied only for some limiting cases ($Bi<<1$ or $Bi>>1$).
In order to determine the critical state stability in an intermediate case, numerical calculations are necessary.
Such calculations were performed e.g. in Ref. \cite{Zhou}.

In order to determine the current-voltage characteristics, we study the relaxation of the magnetic moment.
Although Eq. 2 can be applied to describe the the current-voltage characteristics in a wide range of the electric field,
 one must bear in mind that the electric field induced in the superconducting sample during magnetic relaxation experiments is significantly lower
 than the electric field induced by the external magnetic field sweep.

One should also bear in mind that the accuracy of determination of $H_\mathrm{fj1}$ for the highest sweep rate ($2 \cdot 10^{-2}$ T)
in our experiment was only about  $\pm 0.1$ T.
This experimental error has a special significance for the sample orientation with the external magnetic field parallel to the $ab$-plane,
 because for this orientation all observed values of $H_\mathrm{fj1}$ were in the range between 0.5 T and 1 T (see Fig. 2).

Finally, more accurate analysis of the critical state stability in an anisotropic material should also take into account the value of the out-of-plane thermal conductivity.

Our experimental results, as well as the data for textured YBaCuO \cite{Guillot}, show that by the improvement of the cooling conditions,
 i.e. by immersing the sample in liquid helium, the critical state stability increases significantly.
Both dynamic models used in our analysis predict the dependence of $H_\mathrm{fj1}$ on the external magnetic sweep rate (see Eq. 3 and Eq. 5).
These equations were derived assuming the critical current density to be independent on the magnetic field and predict this dependence
 to be described by a power function with the power of $-\frac{1}{2}$ \ and $-\frac{1}{3}$ \ 
for the approximation of weak cooling and the isothermal approximation, respectively.
However, if the critical current density is strongly dependent on the magnetic field,
 the dependence of $H_\mathrm{fj1}$ on the sweep rate can be modified \cite{Mints96}.
The comparison of the predictions of Eq.~3 with the experimental results, one can see in Fig.~2 and in Fig.~4.
One can observe the largest inconsistency for the lowest temperatures and for the external magnetic field parallel to the $ab$-plane.

The influence of the heat exchange conditions on critical state stability as well as a strong dependence of $H_\mathrm{fj1}$ on the sweep rate show
 that thermomagnetic instabilities are initiated in dynamic conditions, in which the magnetic diffusivity is smaller than the thermal one.
On the other hand, the effective diffusivity describing the dynamics of the flux jumps (see Fig. 5b) is very high,
 it means comparable to the magnetic diffusivity in the normal state.
The magnetic field dependence of this effective diffusivity suggests that it can be correlated with the flux flow resistivity.
In order to understand this discrepancy, one should explain that the initial stage of the avalanche develops at the voltage level
 that is induced by the external magnetic field sweep.
We estimate that the sweep rates used in our experiments induced in the coil the voltage of an order of $10^{-8}-10^{-6}$ V,
 which is six or four orders of the magnitude lower than the noise level registered by the acquisition board.
For this reason, we were not able to register this very initial stage of the avalanche.
Our experimental results show that after this initial stage of the avalanche the resistivity of the sample rapidly increases
 and the dynamics of the flux jumps is governed by the magnetic diffusivity, which depends of the flux flow resistivity.

\section{Conclusions}

The stability of the critical state in a large crystal of \La \ is influenced by its anisotropy,
 and it depends on the sample orientation with respect to the external magnetic field.
The most relevant is the anisotropy of the critical current.
At relatively slow magnetic field sweep rates, thermomagnetic avalanches are initiated in the dynamic conditions.
Hence, one observes a strong influence of the heat exchange conditions as well as of the sweep rate on the critical state stability.
In the case of the experiments, where the investigated sample is surrounded by helium gas with low pressure (PPMS),
 the critical state stability criterion can be derived in the approximation of a weakly cooled sample.
However, this approximation cannot be applied in the case of large crystals immersed in liquid helium.
Flux jumps in the single crystalline \La \ are very short, typical duration is of several microseconds.
The dynamics of the jumps is governed by the magnetic diffusivity, proportional to the flux flow resistivity.

\ack

This work was partly supported by Polish National Science Centre under research project for years 2011-2012 (grant N N202 1663 40).

\end{document}